*Complex Magnetic Phase Diagram of a Geometrically Frustrated Sm Lattice: $SmPd_2Al_3$ case*

J. Pospíšil[1*], G. Nénert[2], S. Miyashita[3], H. Kitazawa[4], Y. Skourski[5], M. Diviš[1], J. Prokleška[1] and V. Sechovský[1]

[1]*Faculty of Mathematics and Physics, Department of Condensed Matter Physics, Charles University, Ke Karlovu 5, 121 16 Prague 2, The Czech Republic*
[2]*Institut Laue Langevin, BP 156, 6 rue Jules Horowitz, 38042, Grenoble Cedex 9, France*
[3]*Department of Physics, School of Science, The University of Tokyo 7-3-1 Hongo, Bunkyo-ku, Tokyo, 113-0033 Japan*
[4]*National Institute for Materials Science, Tsukuba, Ibaraki 305-0047, Japan*
[5]*Dresden High Magnetic Field Laboratory, Helmholtz-Zentrum Dresden Rossendorf, D-01314 Dresden, Germany*

**Abstract**

Magnetism in $SmPd_2Al_3$ was investigated on a single crystal by magnetometry and neutron diffraction. $SmPd_2Al_3$ represents a distinctive example of the Sm magnetism exhibiting complex magnetic behavior at low temperatures with four consecutive magnetic phase transitions at 3.4, 3.9, 4.3 and 12.5 K. The rich magnetic phase diagram of this compound reflects the specific features of the $Sm^{3+}$ ion, namely the energy nearness of the ground-state multiplet $J = 5/2$ and the first excited multiplet $J = 7/2$ in conjunction with strong crystal field influence. Consequently, a significantly reduced Sm magnetic moment in comparison with the theoretical $Sm^{3+}$ free-ion value is observed. Despite the strong neutron absorption by natural samarium and the small Sm magnetic moment ($\sim 0.2\ \mu_B$) we have successfully determined the magnetic k-vector (1/3, 1/3, 0) of the phase existing in the temperature interval 12.5 - 4.3 K. This observation classifies the $SmPd_2Al_3$ compound as a magnetically frustrated system. The complex magnetic behavior of this material is further illustrated by kinetic effects of the magnetization inducing rather complicated magnetic structure with various metastable states.

## Introduction

The SmPd$_2$Al$_3$ compound belongs to the class of the rare earth materials crystallizing in the hexagonal crystal structure of the PrNi$_2$Al$_3$-type (space group P6/mmm)[1]. The physical properties of the rare earth counterparts with the composition REPd$_2$Al$_3$ (for RE= Ce, Pr, Nd Sm and Gd) are given mainly by the strong influence of the crystal field on the magnetic state of RE ions[1,2]. This leads to various types of magnetic order (Ce, Nd, Sm and Gd)[3-7] or contrary paramagnetic ground state (Pr)[8] and heavy fermion behavior (Ce).[9] Finally, the Y and La compounds are superconductors[10-12]. The REPd$_2$Al$_3$ compounds are therefore an interesting playground for theoreticians as they are modeling examples to study rare earth elements magnetism due to their high variability of the physical properties connected with a simple and high symmetry crystal structure.[10, 13-15] Although physical properties of all compounds in the REPd$_2$Al$_3$ series were subjected to intensive research activities, yet the two most interesting cases - Gd and Sm compounds remain poorly understood despite the recent progress within the last two years.[13, 16]

SmPd$_2$Al$_3$ was described as an antiferromagnet with strong uniaxial anisotropy even in the paramagnetic state with an easy-magnetization direction along the crystallographic axis $c$. Four successive magnetic transitions have been identified in the temperature dependence of the specific heat at temperatures $T_3 = 3.4$ K, $T_2 = 3.9$ K, $T_1 = 4.3$ K and $T_C = 12.5$ K. The high number of the magnetic phase transitions and the series of four magnetic field induced transitions detected at 0.03, 0.35, 0.5, and 0.75 T, respectively, at 1.8 K yield a complex magnetic phase diagram.

Generally, the complexity of the Sm magnetism is given by anomalous magnetic ground state of the Sm$^{3+}$ ion.[17] The Sm$^{3+}$ ground-state multiplet $J = 5/2$ is radically influenced by first and second excited multiplets $J = 7/2$ and $J = 9/2$ which have only an energy of 0.1293 eV and 0.2779 eV[17] which results in distinctive features like multiple magnetic phase transitions and the susceptibility influenced via the temperature-independent Van Vleck term.[18]

Neutron scattering is usually a good tool to study magnetism on a microscopic scale. However, Sm containing materials are usually disregarded due to high thermal-neutron absorption by of natural samarium. In addition to the absorption problem, another difficulty arises from the usually low magnetic moment of Sm$^{3+}$. The natural Sm consists of 7 isotopes with abundances shown in the Table 1.[19, 20]

TABLE I. Table summarized information about all isotopes presented in natural Samarium-their atomic masses, natural abundance and their neutron absorption.

| Isotope | Atomic mass (ma/u) | Natural abundance (atom. %) | Absorption (barn) |
|---|---|---|---|
| $^{144}$Sm | 143.911998 (4) | 3.07 (7) | 0.7 |
| $^{147}$Sm | 146.914894 (4) | 14.99 (18) | 57 |
| $^{148}$Sm | 147.914819 (4) | 11.24 (10) | 2.4 |
| $^{149}$Sm | 148.917180 (4) | 13.82 (7) | 42080 |
| $^{150}$Sm | 149.917273 (4) | 7.38 (1) | 104 |
| $^{152}$Sm | 151.919728 (4) | 26.75 (16) | 206 |
| $^{154}$Sm | 153.922205 (4) | 22.75 (29) | 8.4 |

The high thermal-neutron absorption of the natural samarium ($^{nat}$Sm) is given mainly by isotopes $^{149}$Sm, $^{150}$Sm and $^{152}$Sm (see Table I) with total average absorption of the $^{nat}$Sm 5922 barn[21-23]. In order to overcome the problem of absorption, there are 2 possibilities. The first one is to work with isotopic samarium, typically $^{154}$Sm which combines low neutron absorption and high coherent scattering length. Unfortunately the cost of $^{154}$Sm isotope (99%) metal is prohibitive. The second choice is to make use of the opportunity that the magnitude of the neutron absorption can strongly depends on neutron energy[24]. Consequently, higher energy neutrons ("hot" neutrons[25, 26]) are a good alternative. Therefore, we have carried out neutron single crystal experiment using the D9 high resolution diffractometer at hot source in the Institut Laue Langevin, Grenoble, France.

In this paper, we construct a detailed magnetic phase diagram of SmPd$_2$Al$_3$ using SQUID magnetometry and investigate the nature of the first magnetic phase using single crystal neutron diffraction on natural isotopic SmPd$_2$Al$_3$.

**Experimental and computational details**

The single crystal of the SmPd$_2$Al$_3$ compound has been grown in a triarc furnace by Czochralski pulling method from stoichiometric amounts of elements. Pulling and single crystal growth details and quality were already described in Ref.[13]. The natural isotope of Sm has been used.

The samples of the appropriate shape for the magnetization and neutron experiment have been cut by a wire saw (South Bay Technology Inc, type 810). The sample for the magnetization measurements had the following dimensions: 1x1x1.5 mm$^3$ with rectangular planes oriented perpendicular to the crystallographic axes *a* and *c*. A single crystal of size 1.9 x 1.8 x 2.2 mm$^3$ was used for the neutron diffraction experiment. All the planes of the samples were gradually polished and clean using 6, 3 and 1-micron diamond particles suspension. The orientation of each as-prepared sample was checked by backscattering Laue technique using Cu white X-ray radiation before measurements.

The magnetization measurements were performed using a commercial Quantum Design MPMS (Magnetic Property Measurement System) device. To calculate magnetic isotherms, we have employed crystal field model introduced in our previous work.[13] The high-field magnetization experiment was carried out with the extraction method using the 40 T class hybrid magnet in High Magnetic Field Laboratories of the National Institute for Materials Sciences in Tsukuba, Japan. The pulse magnetic field experiment was realized in 60 T magnet in Dresden High Magnetic Field Laboratory (HZDR) in Germany.

Single-crystal neutron diffraction data were collected on the high resolution four-circle diffractometer D9 at the Institut Laue-Langevin, Grenoble, using a wavelength of 0.5109(1) Å obtained by reflection from a Cu(220) monochromator. The wavelength was calibrated using a germanium single crystal. D9 is equipped with a small two-dimensional area detector,[27] which for this measurement allowed optimal delineation of the peak from the background. For all data, background corrections following Wilkinson et al. (1988) and Lorentz corrections were applied.[28]

**Crystal structure analysis**

Using single crystal neutron data we have confirmed that SmPd$_2$Al$_3$ crystallizes in the space group P6/mmm with cell parameters *a* = *b* = 5.3970(4) Å and c = 4.1987(5) Å at 20 K. There are very similar to previously published cell parameters.[1, 7] No crystal structure phase transition has been observed from room temperature down to 2 K. The crystal structure refinement at 20 K was based on the data collection of 149 unique reflections. The data were corrected for absorption. The best data fit with $R(F^2) = 6.14\%$ is presented in Fig. 1. The corresponding atomic coordinates which are all in special Wyckoff positions and the anisotropic displacement parameters are given in Table II.

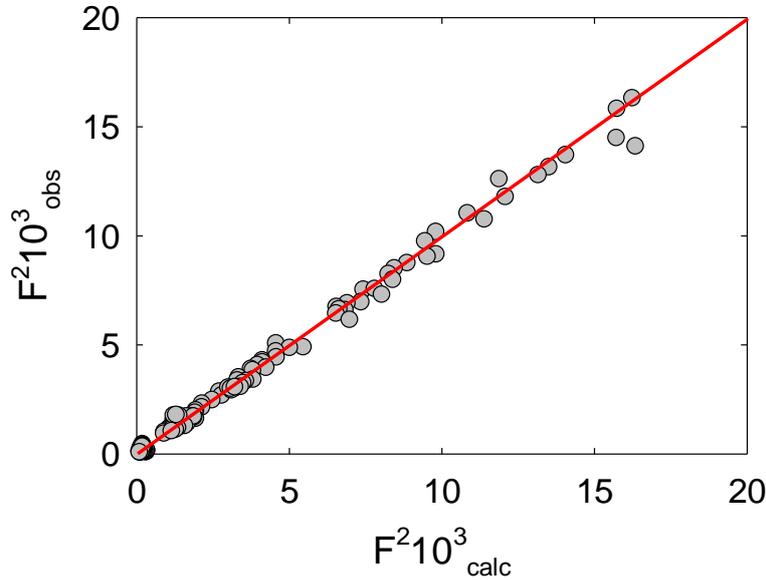

FIG. 1. Calculated versus observed of the squared intensities for the data collected at 20 K.

TABLE II. Atomic coordinates and the anisotropic displacement parameters as determined at temperature 20 K.

|    | X    | Y    | Z   | $U_{11}$   | $U_{22}$   | $U_{33}$    | $U_{12}$   | $U_{13}$ | $U_{23}$ |
|----|------|------|-----|------------|------------|-------------|------------|----------|----------|
| Al | 1/2  | 0(-) | ½   | 0.0028(7)  | 0.0021(7)  | 0.0028(9)   | 0.0011(7)  | 0(-)     | 0(-)     |
| Pd | 1/3  | 2/3  | 0(-)| 0.0012(6)  | 0.0012(6)  | 0.0007(6)   | 0.0006(6)  | 0(-)     | 0(-)     |
| Sm | 0(-) | 0(-) | 0(-)| 0.0081(8)  | 0.0081(8)  | 0.0084(11)  | 0.0040(8)  | 0(-)     | 0(-)     |

**Magnetic phase diagram study**

First, we studied magnetization curves at low temperature $T = 1.7$ K and magnetic field up to 30 T using a hybrid magnet (at Tsukuba Magnet Laboratory) with field applied along the crystallographic axis *c*, *a* and the in-plane *210* direction. We have found clear evidence of the easy-axis type anisotropy with axis *c* as the direction of easy magnetization (see Fig. 2). If one assumes the constant slope of the magnetization curve for each axis without any metamagnetic transition, the magnetization along the hard axis in the *ab* plane will attain to the magnetization along the *c* axis at 79.8 T.

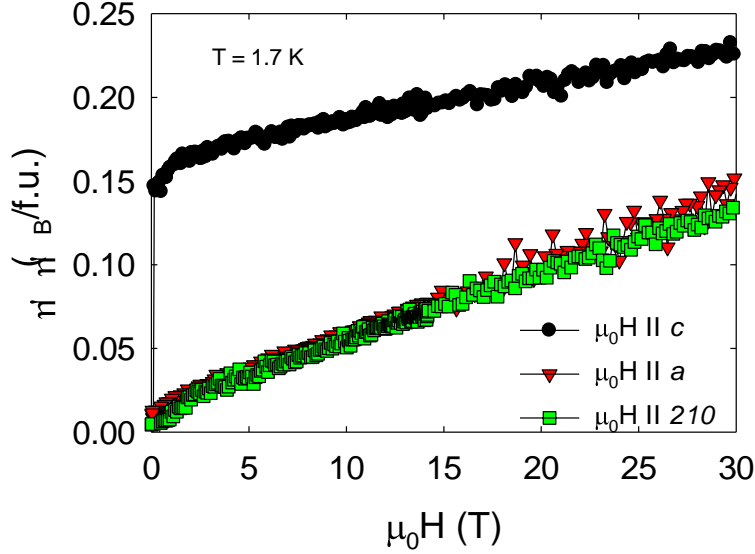

FIG. 2. Magnetization curves measured with magnetic field applied along the *a*, *c* crystallographic axis and also in the in-plane *210* direction at 1.7 K.

The behavior of the magnetization curves clearly denotes a strong easy-axis-type magnetic anisotropy. A small discrepancy of magnetization curves for the hybrid magnet and the SQUID magnetometer along the c-axis (discussed later) may come from the difference of the temperatures of measurements and/or a wrong estimation of background.

Before exploring the temperature and magnetic field dependence of the phase diagram in more details, we have focused on the magnetic behavior of the $SmPd_2Al_3$ compound published in previous works. Precise magnetization loops were measured exhibiting complex magnetic features with complicated step-like shape.[6, 13]

The heat capacity data published in Ref.[13] show four successive magnetic transitions at $T_3 = 3.4$ K, $T_2 = 3.9$ K, $T_1 = 4.4$ K and $T_C = 12.5$ K in the zero magnetic field. From the published crystal field analysis, we may regard the system as a spin $S = 1/2$ system. The energy gap to the next spin doublet is about 100 K and it would not affect low temperature properties below 12 K. The spin degree of freedom exhibits magnetic phase transitions (see Ref.[13]).

Finally, the magnetic phase diagram of the $SmPd_2Al_3$ is complicated not only due to the four successive magnetic transitions observed in the specific heat data in zero magnetic field but furthermore by a series of field-induced magnetic transitions (see Refs.[6, 13]).

On the ground of the heat capacity data (Ref.[13]), we have precisely investigated the temperature evolution of the magnetization loops by 0.1 K temperature steps in the temperature range 1.8 – 5 K while applying the magnetic field along the *c* axis. Another four magnetization loops have been measured also in the temperature range 5 – 12 K. We present these magnetization loops in Figs. 3, 4 and 5.

In Fig. 3a, we present three representative loops for the interval between 1.9 and 2.9 K which covers the region below $T_3$. Three magnetic field-induced transitions can be identified in the magnetization curves within this interval. In Figs. 3b and 3c, we present a zoom into the magnetic field regions where the phase transitions occur. The onset of the first transition is located at $\mu_0H_{C1} = 0.038$ T. We call the phase defined by $\mu_0H < \mu_0H_{C1}$ as *P1*. The second field-induced transition is smooth. It starts to appear at a magnetic field of 0.35 T and is completed at 0.55 T where it terminates. The onset of 0.55 T will be considered as a point for construction of the magnetic phase diagram. We will call this phase as *P2*. The determination of the third transition is rather tedious as it is quite broad. For simplicity, we will take into account the upper onset of 2.5 T that is also the point used for construction of the phase diagram. The third phase will be called as *P3*. To study the hysteresis our attention has been also focused on the second branches of the loops in the magnetic field decreasing from 5 T to 0 T. The first drop of the magnetization has been found in the magnetic field of 1.9 T which denotes the hysteresis of 0.6 T for the magnetic phase *P3* (area of the hysteresis is marked as *P3a* - Fig. 6). The first drop of magnetization is followed by continuous decrease of the magnetic moment, which ends in the magnetic field of –0.02 T by magnetization reversal. It means the hysteresis of the *P1* phase of 0.018 T (area of the hysteresis is marked as *P1a* - Fig. 6). The subsequent temperature increase from initial 1.9 K leads to reduction of the hysteresis both phases *P1a* and *P3a* and all field induced transitions are shifted to lower magnetic field as it is marked by arrows in the panel B) of Fig. 3.

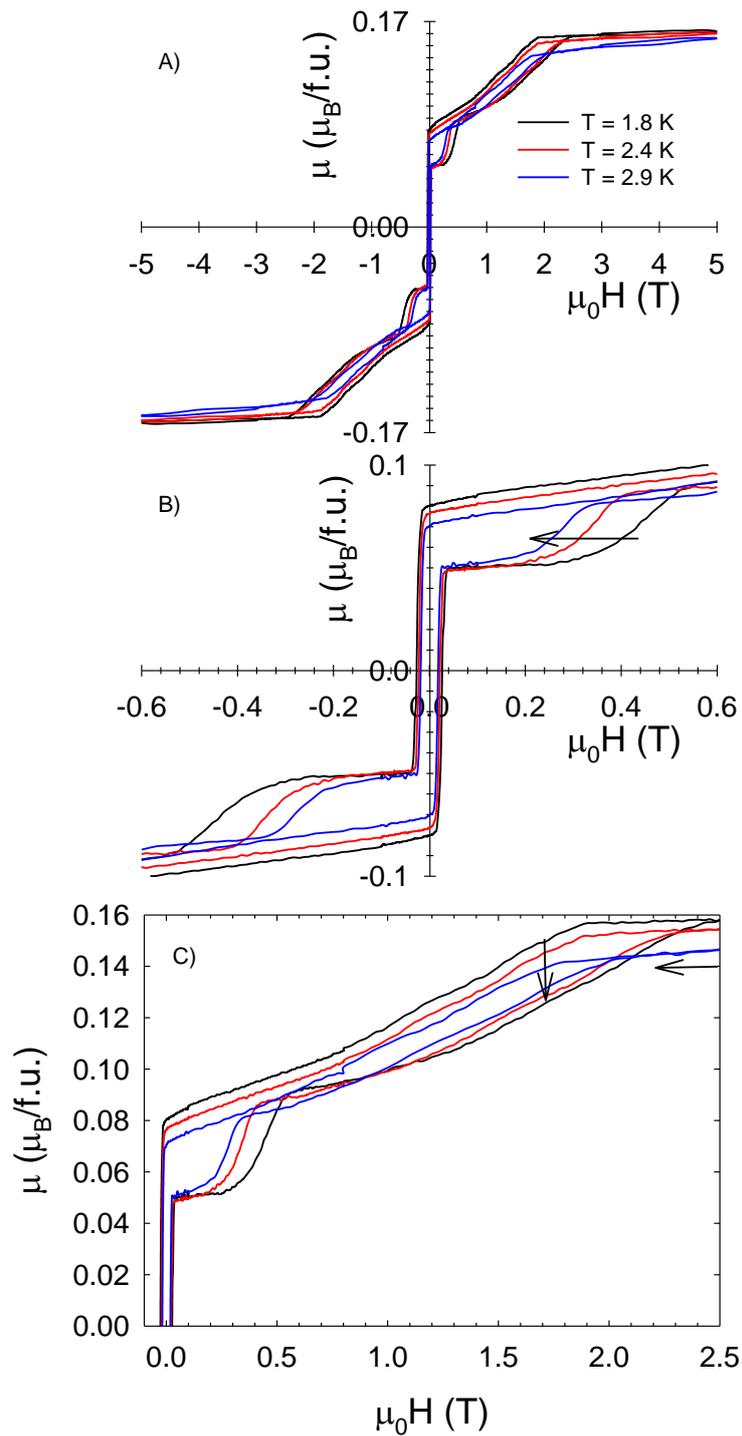

FIG. 3. A) Magnetization loops measured in the temperature interval between 1.8 – 2.9 K. Only three representative loops are displayed. B) Low magnetic field features of the loops. Gradual vanishing of the hysteresis and shifts of the jumps in the loops to lower magnetic fields is indicated by arrows. C) Magnetic field loops in the positive field quadrant up to the saturation field.

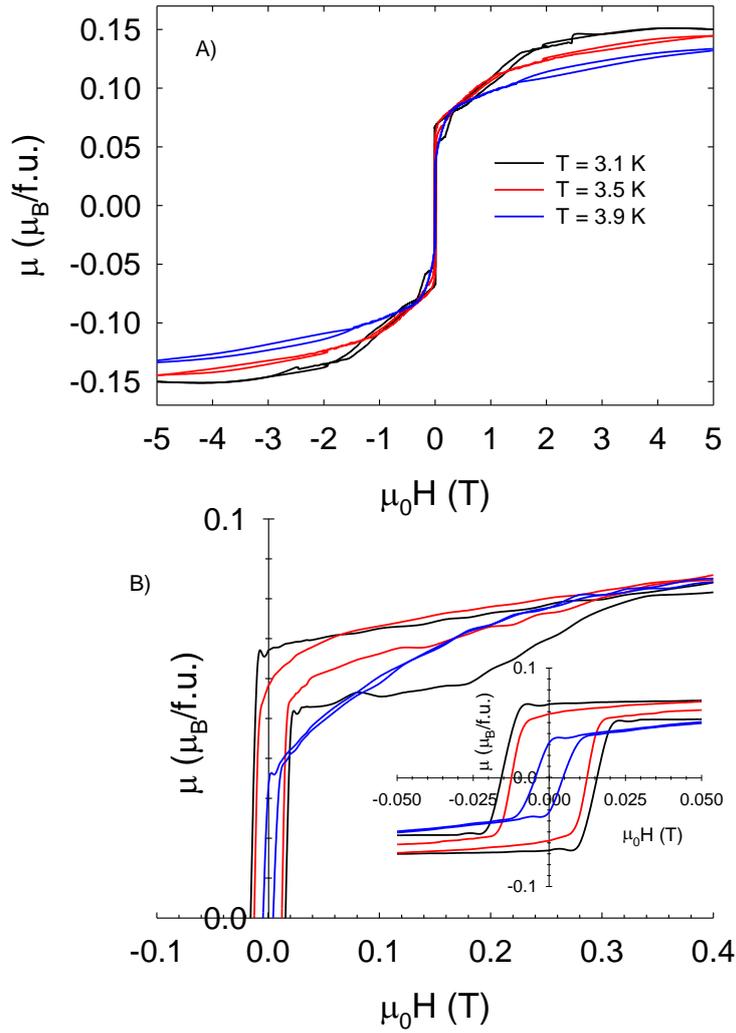

FIG. 4. A) The representative of magnetization loops measured in the temperature interval between 3.1 – 3.9 K. B) Low magnetic field features of the loops. Gradual vanishing of the hysteresis and shifts of the jumps in the loops to lower field. Dramatic changes occurs in this temperature interval when hysteresis (*P3a*) disappears at ~ 3.5 K and also the jump of the *P2* phase disappears at a temperature between 3.7- 3.9 K.

The other set of figures (Figs. 4a and 4b) represents the temperature evolution of the hysteresis loops in the temperature interval between 3.1 – 3.9 K. The first significant change of the loops is expected in this interval because two magnetic transitions were predicted at temperatures 3.4 and 3.9 K in specific heat data.[13] All magnetic field induced transitions described in Fig. 3 are conserved up to 3.4 K but dramatic suppression of the transition (jump) originally occurring in field interval 0.35 T - 0.55 T (marked as *P2* phase) is evident. It is followed by fast reduction of the hysteresis of the upper phase marked as *P3a* in the later constructed phase diagram. The critical point appears at a temperature of ~3.5 K where the hysteresis of the upper phase *P3a* disappears. The second dramatic change is merging of the phases *P2* and *P3* at a temperature between 3.7- 3.9 K. Above the temperature of ~ 3.8 K only the hysteresis of the *P1a* phase appears in the low magnetic field and the common weak knee of the phases *P2* and *P3* appearing around magnetic field 0.2 - 0.3 T is conserved (Fig. 4). The new phase rising from the original *P2* and *P3* phases is newly called *P4*.

The third set of the loops (Figs. 5a and 5b) comes from the temperature interval between 4.1 – 4.9 K. This temperature interval is characterized by erase of the last of the phases *P1*

(including its hysteresis *P1a*) in the low magnetic fields. The hysteresis and also *P1* phase simultaneously disappear at a temperature of ~4.5 K (Fig. 5 panel B). The temperature interval between temperatures 3.8 – 4.4 K is characterized by coexistence of the *P4* and *P1* phase with *P1a* hysteresis. Above 4.4 K only the *P4* phase survives as a weak knee up to ~12 K where the loop has a paramagnetic like shape.

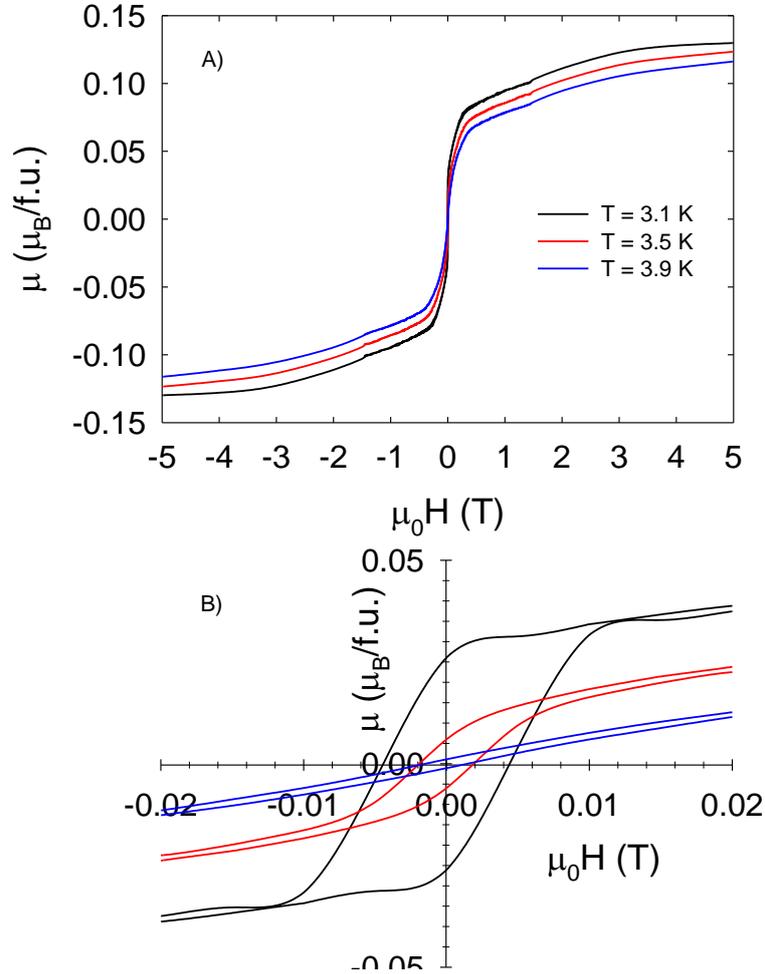

FIG. 5. A) The group of magnetization loops measured in the temperature interval between 4.1 – 4.9 K. B) Low magnetic field features of the loops. Gradual vanishing of the hysteresis *P1a* and *P1* phase is evident. Dramatic change occurs when the hysteresis (P1a) and P1 phase simultaneously disappear at the temperature of ~4.5 K and only the *P4* phase survives up to ~12 K.

Taking into account results of the magnetization measurements, we have finally constructed the phase diagram of $SmPd_2Al_3$ compound (see Fig. 6).

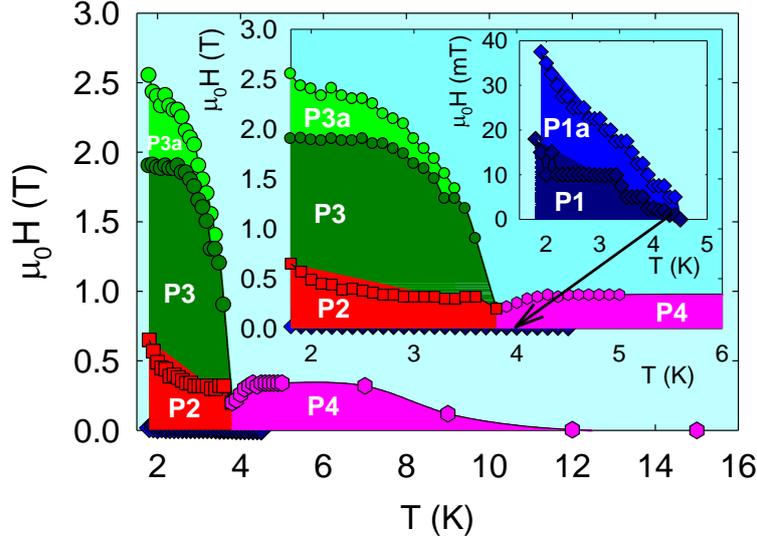

FIG. 6. Magnetic phase diagram of the $SmPd_2Al_3$ compound constructed on the basis of magnetization data in the magnetic field applied along the c-axis.

**Magnetic structures study**

Based on the magnetization data, we have sketched the complex magnetic phase diagram of $SmPd_2Al_3$ as shown in Fig. 6. However, the nature of the various magnetic phases remained unknown. Consequently, we have carried out single crystal neutron diffraction experiment using the D9 four cycle diffractometer with a short wavelength $\lambda = 0.511$ Å. We have cooled down the crystal below $T_C$ and carried out $q$-scans. Despite a weak magnetic signal (the saturated magnetic moment is only of 0.16 $\mu_B$/f.u.) and the strong absorption, we were able to observe magnetic reflections. One of the strongest magnetic reflections (5/3 5/3 0) is illustrated in Fig. 7. The Miller indices of the magnetic reflections are of the type (h k 0) in good agreement with magnetization data where the *c*-axis represents the easy magnetization axis; the k-vector of the ground state magnetic structure has been determined as **k** = (1/3, 1/3, 0) and symmetry conditions given by P6/mmm space group. No component of the magnetic moment has been detected in other crystallographic directions by magnetization measurements.

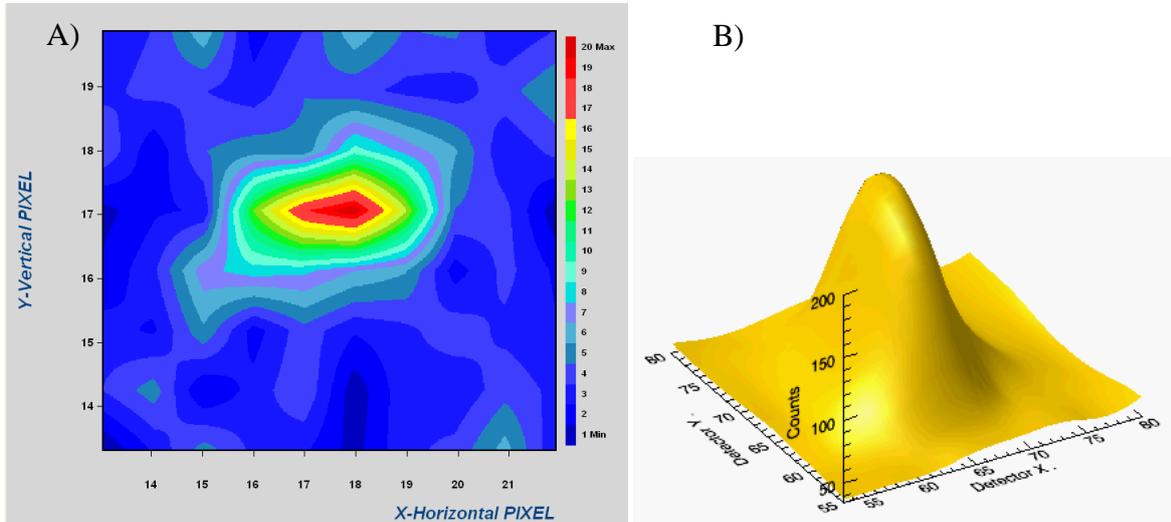

FIG. 7. Reflection (5/3 5/3 0) as observed at 3.6 K. A) Filled contour plot showing the reflection as recorded in the 2D detector. B) 3D plot of the reflection.

We could do acquisition at the top of a restricted number of magnetic reflections using the 2D area detector. Integrated intensities as functions of temperature using omega scans could not be determined due to the weakness of the magnetic reflections. We followed few reflections as functions of temperature. We present in Fig. 8 the temperature dependence of the (5/3 5/3 0) reflection which appeared to be the strongest.

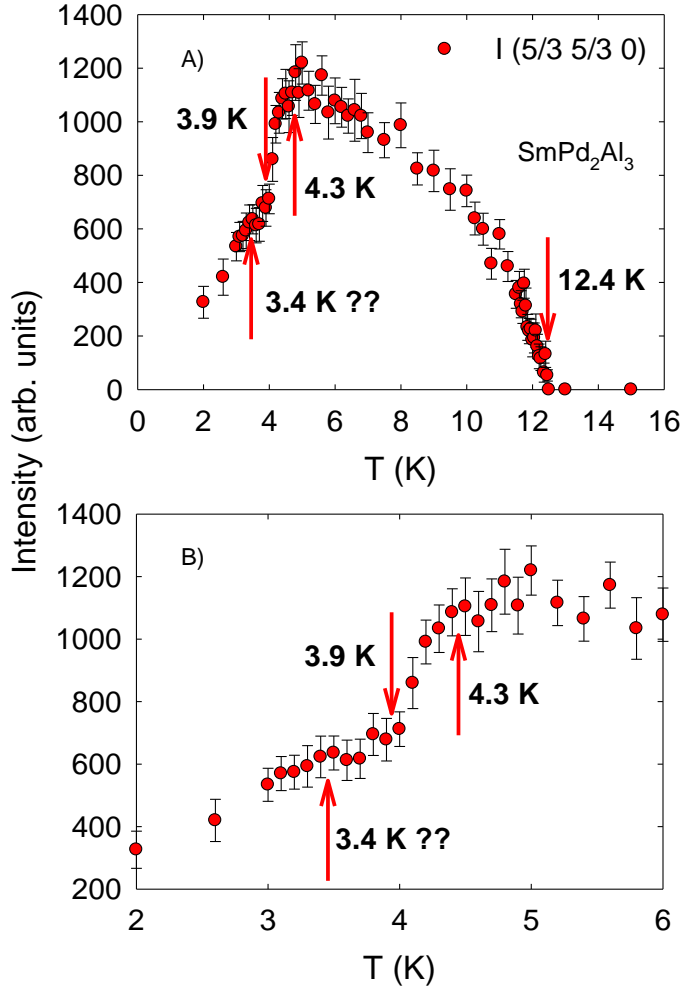

FIG. 8. Temperature dependence of the (5/3 5/3 0) reflection. In A), we show the temperature evolution in the whole temperature range. B) Zoom in the temperature range 2 to 6K. The red arrows mark the transition temperatures as determined from specific heat. The presence of the last transition at temperature $T_3 = 3.4$ K is disputable within the error bars.

We can clearly see that the magnetic reflection (5/3 5/3 0) emerges at about 12.4 K, which well corresponds to the critical temperature $T_C$ determined from specific heat measurements.[4-7, 13] Further decrease of temperature leads to an increase of the (5/3 5/3 0) reflection intensity with a maximum around 4.5 - 5 K which quite well corresponds to $T_1$.[13] When further decreasing temperature the intensity of the (5/3 5/3 0) reflection shows a drop around 4 K. This corresponds to the critical temperature $T_2$.[13] The expected anomalies at $T_1$ and $T_3$ are not so clearly visible. The presence of the last anomaly $T_3$ at temperature 3.4 K is open question when taking into account the error bars of the intensity. The sudden drop of the intensity at around 4.5 K is probably related to a change of the propagation vector as it goes away. However, the extremely weak magnetic signal and the lack of resolution in q due to short neutron wavelength prevented confirmation of this hypothesis.

**Discussion**

Below $T_C$, the intensity of the reflection (5/3 5/3 0) behaves as an order parameter which could be fitted to a power law as $I = a * (T_C - T)^\beta$. The resulting fit is presented in Fig. 9. The obtained critical exponent is close to 0.5 suggesting that the behavior of SmPd$_2$Al$_3$ between $T_C$ and $T_1$ can be understood within the mean field theory.

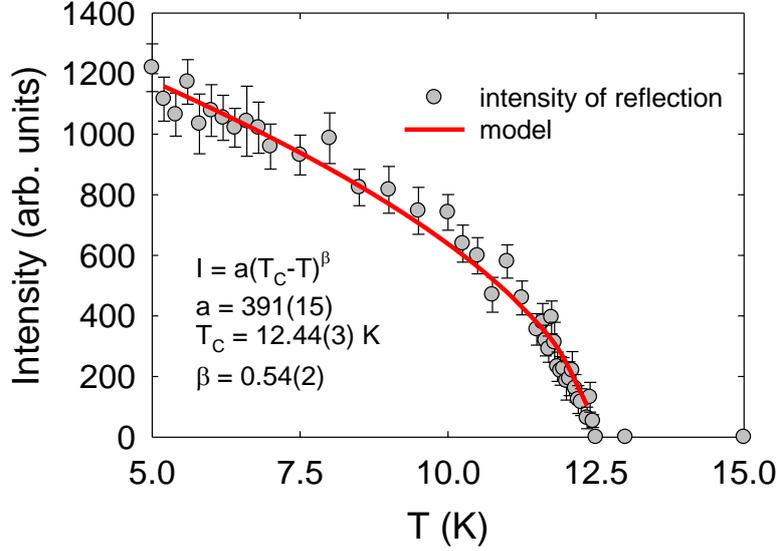

FIG. 9. Fit of the intensity of the reflection (5/3 5/3 0) as function of temperature.

In addition, we have inspected several other measured magnetic reflections such as (4/3 1/3 0), (1/3 4/3 0), (1/3 1/3 0) and (2/3 2/3 0). No magnetic reflection with l ≠ 0 has been detected. This observation is consistent with the magnetic ordering of Sm magnetic moments parallel to the *c* axis. Nevertheless, we are aware that a possible slight off the c-axis component cannot give magnetic reflections detectable within our experiment and therefore we can take the scenario with the Sm magnetic moments parallel to the c-axis only tentatively. Taking into account that the propagation vector between $T_C$ and $T_1$ is **k** = (1/3, 1/3, 0) and there is a priori no magnetic component in *ab* plane, we can give a representation of this likely magnetic structure (see Fig. 10). The magnetic unit-cell is 3 times larger along *a* and *b*, respectively, and is formed by 2 hexagonal sublattices which are interpenetrated and are coupled antiferromagnetically. The coupling along the *c*-axis is ferromagnetic.

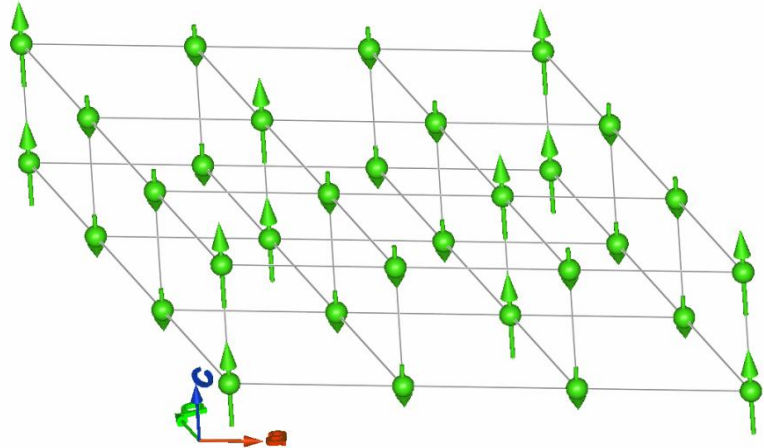

FIG. 10. Likely magnetic structure of the phase stable between $T_C$ and $T_1$ in SmPd$_2$Al$_3$.

The magnetic structure between $T_C$ and $T_1$ characterized by **k** = (1/3, 1/3, 0) with a hexagonal lattice can be discussed in the scenario of magnetic frustration. As the model example the isostructural compound $GdPd_2Al_3$ can be considered being presented as a magnetically frustrated Heisenberg triangular lattice antiferromagnet with weak Ising anisotropy.[16, 29, 30] This can be a key for understanding of the magnetic structures of the phases in the phase diagram of the related $SmPd_2Al_3$. Both compounds, $GdPd_2Al_3$ and $SmPd_2Al_3$, have many similarities but also some different magnetic features. The main difference between the Sm and Gd compounds comes from the magnetic state and influence of the crystal field on magnetic ions. The $Gd^{3+}$ ion represents an exception among rare earth ions because of its zero angular momentum. Due to this fact the multiplet $J = {}^8S_{7/2}$ ground state remains fully degenerate in the crystal field. In the absence of external magnetic field, only an exchange magnetic field can lift the $(2J + 1)$-fold degeneracy.[31, 32] The $Sm^{3+}$ ion represents a totally different case as it was suggested in the Introduction. The connecting point between the two compounds is the same hexagonal crystal structure with magnetic **k**-vector (1/3, 1/3, 0) (temperature interval $T_C - T_1$) for $SmPd_2Al_3$ and also for $GdPd_2Al_3$ between $T_{N1}$ and $T_{N2}$.[16]

Generally, the presence of magnetic frustration in solids is revealed by a few experimental evidences in first simple approach. The first of them is the existence of plateaus in magnetization curves[33-35] and anomalously low $\theta_{CW}$ with respect to critical temperatures[36, 37]. The empirical quantity $f = -\theta_{CW}/T_C > 1$ corresponds to frustration (2 for antiferromagnets). The investigation of both parameters is pretty complicated for the $Sm^{3+}$ magnetic state. The reciprocal susceptibility $1/\chi$ of $SmPd_2Al_3$, which is affected by temperature-independent Van Vleck contribution due to the low-lying first excited multiplet $J = 7/2$ being populated[13] does not follow the Curie-Weiss law. Despite of this fact the fitted $\theta_{CW} = -21.3$ K using a modified Curie-Weiss law[1] and our found $T_C = 12.5$ K gives $f = 1.7$ which is close to expected value for frustrated antiferromagnets.[37] The analysis of the magnetization plateaus is not straightforward due to the missing saturated moment of the $Sm^{3+}$ ion. For comparison, the isostructural compound $GdPd_2Al_3$ is characterized by the well-defined wide 1/3 plateau on the magnetization curve in the magnetic field interval between 6.2 and 11.8 T.[29, 30] Such behavior is typical for triangular lattice antiferromagnets with weak Ising like anisotropy.

The saturated magnetization value of 0.16 $\mu_B$/f.u. deduced from the magnetization data is significantly less than the expected magnetic moment of $gJ\mu_B = 0.71$ $\mu_B$ for the $Sm^{3+}$ free ion. This 0.16 $\mu_B$/f.u. value of the saturated magnetic moment is comparable with value found in Ref.[13]. This considerably reduced saturated magnetic moment value by a factor of about five motivated us to carry out high magnetic field experiment up to 60 T in the pulse field magnet. The pulse field was applied along the easy magnetization $c$ axis. Our high magnetic field experiment up to 60 T did not show any additional features compared to our lower magnetic field measurement. Especially no extra magnetic field induced phases were observed either along the $c$ axis or in $ab$ plane.

In order to gain some more insight into our system, we have carried out theoretical calculations. To calculate the magnetic isotherms we have employed a crystal field model. The microscopic crystal field Hamiltonian has the hexagonal symmetry and reasonable crystal field parameters were found by first principles calculations in our previous work[13]. The total (CF + Zeeman) Hamiltonian has been diagonalized and the obtained eigen-values and eigenvectors has been used to calculate magnetic isotherms along the $c$- and $a$- axes, respectively. The result of the calculation is presented in Fig. 11.

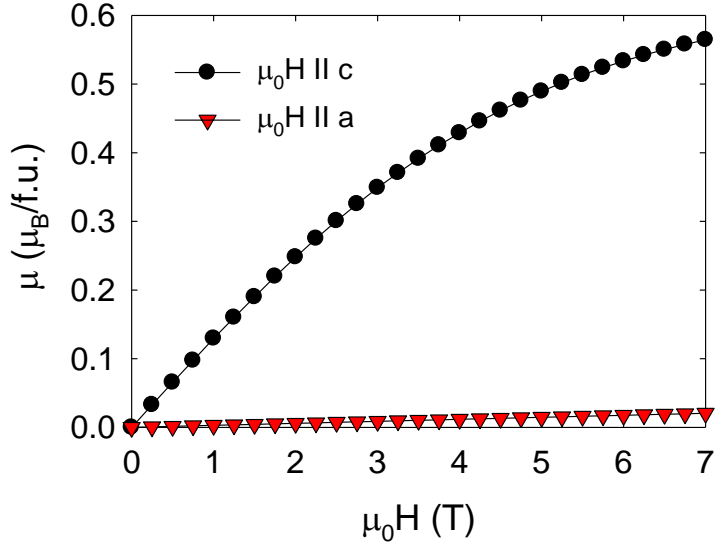

FIG. 11. Figure represents magnetization isotherms calculated using crystal field model.

Firstly the calculation confirms the *c*-axis as the easy magnetization axis but the evolution of the saturated moment is not in reasonable agreement with the experimental data when significantly higher moment has been found – almost three times higher than experimental result in the magnetic field of 5 T. Based on our model, the theoretical field required to reach the saturated magnetic moment is 220 T. However such magnetic field is not experimentally routinely available nowadays. On the basis of these calculations and experimentally available magnetization data, we still cannot exclude any additional field induced transition in magnetic fields higher than 60 T and the question regarding the value of the saturated magnetic moment is still open.

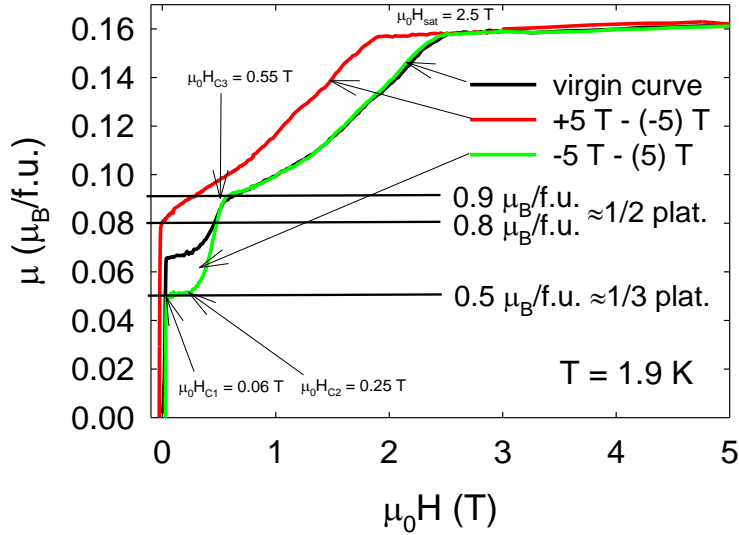

FIG. 12. First quadrant of the magnetization loop of the SmPd$_2$Al$_3$ measured at temperature 1.9 K.

Presently, we adopt 0.16 $\mu_B$/f.u. as the saturated value (Fig. 12). Then, we find, in the field-increasing process, a plateau around 1/3 of the saturated magnetization, and a jump to another plateau around 1/2 of the saturated magnetization. The magnetization gradually increases to the saturated magnetization. On the other hand, in the field-decreasing process, the magnetization decreases to 1/2 of the saturated magnetization. Around zero magnetic field the magnetization shows sharp change (almost jumps) to the opposite sign. But it shows a small hysteresis with $\mu_0 H \approx 0.01$ T. The 1/3 plateau reminds us the magnetization process of antiferromagnets in the triangular (hexagonal) systems. We may also regard the plateau of 0.16 $\mu_B$ as the 1/3 plateau and then we expect another jump to the saturated magnetization at higher field. However, the 1/2 plateau does not fit to this picture. Therefore we cannot take this scenario. Some intermediate kinetic effects could cause the 1/2 plateau in hexagonal systems.[38] Here similar kinetic effect is expected for the ordered state. This type of plateau in the increasing field process has been discussed as Magnetic Foehn effect[39, 40] in single molecular magnets, which occurs as a kinetic effect. The dependence on the time scale will be studied later.

According to the simple triangular scenario of XXZ antiferromagnetic model on the hexagonal lattice, the magnetization at $\mu_0 H$ is given by Ref [33]. For the model of anisotropic coupling

$$H_A = \sum_{\langle ij \rangle} \left\{ J\left(S_i^X S_j^X + S_i^y S_j^y\right) + J_z S_i^z S_j^z \right\} - H_z \sum_z S_i^z$$

the magnetic field at the beginning of the 1/3 plateau is $\mu_0 H_{C1} = 3J$, and the end of the 1/3 plateau is

$$H_{C2} = 3J \frac{(2A - 1 + \sqrt{4A^2 + 4A - 7})}{2}$$

with $A = J_z/J$. The magnetization at $\mu_0 H = 0$ T is given by

$$M_0 = \frac{(A - 1)}{(A + 1)}$$

From the observation we find $M_0 \cong 1/2$. Thus we estimate $A \cong 3$. From the value of $\mu_0 H_{C1}$ in the experiment, we estimate $J = 2T/3$. From these values, $\mu_0 H_{C2}$ is estimated as

$$H_{C2} = 2T(5 + \sqrt{41}) \approx 22T$$

Thus, we may expect another jump around $\mu_0 H = 22$ T. But we did not find this jump in Fig. 2.

Now we consider the shape of magnetization loop from the view point of the kinetic effect. As we saw in the previous section, the magnetization curve of the SmPd$_2$Al$_3$ at temperature T = 1.9 K is characterized by two types of steps (Fig. 12). The first one appears as a jump to a plateau of 1/3 of the saturated moment at small value of the magnetic field. The plateau exists in the magnetic field interval $\mu_0 H_{C1} = 0.045$ T and $\mu_0 H_{C2} = 0.25$ T. Then the magnetization increases and reaches 1/2 of saturated moment to full saturated value at $\mu_0 H_{C3} = 0.55$ T. Then the magnetization curve has a kink or a small plateau and then, the magnetization gradually increases to full saturated value. It reaches the saturated moment at $\mu_0 H_{C4} = 2.5$ T. In the process of decreasing magnetic field, the magnetization begins to decrease from the saturated value at $\mu_0 H_{C5} = 1.9$ T and decreases gradually to near 1/2 of saturated moment and jump to the negative value around $\mu_0 H = 0$ T. Because of the hysteresis of the magnetization process, the magnetization curve cannot be considered in truth as a plateau.

There is also necessary to consider other effects that can lead to induction of the plateaus and jumps in magnetization curves. Dynamical magnetic processes have been found in single molecular magnets and also magnetic rings. For example, a phonon-bottleneck effect in $V_{15}$[41] which was explained as a phenomenon due to lack of photon mode for the equilibration. Similar phenomenon was observed in [Fe(salen)Cl]$_2$[42] and also Fe$_{10}$[43]. These phenomena can be regarded as adiabatic processes with small inflow of heat. Such effect of the influence of kinetic effect of sweeping magnetic field is understood as magnetic Foehn effect.[34,35] Similar phenomena have been observed in macroscopic magnetization processes. For example, Katsumata et al. has found it in FeCl$_2$.2H$_2$O[38] and also Narumi et al. in Kagome lattice[44].

In the present case the step is found in a macroscopic change of magnetization. To clarify the influence of the sweeping magnetic field on magnetization we have measured magnetization loop both with a slow field rate in SQUID magnetometer and in pulse field magnet where the maximum used field of 8 T was reached within few tens milliseconds. The results are shown in the Fig. 13.

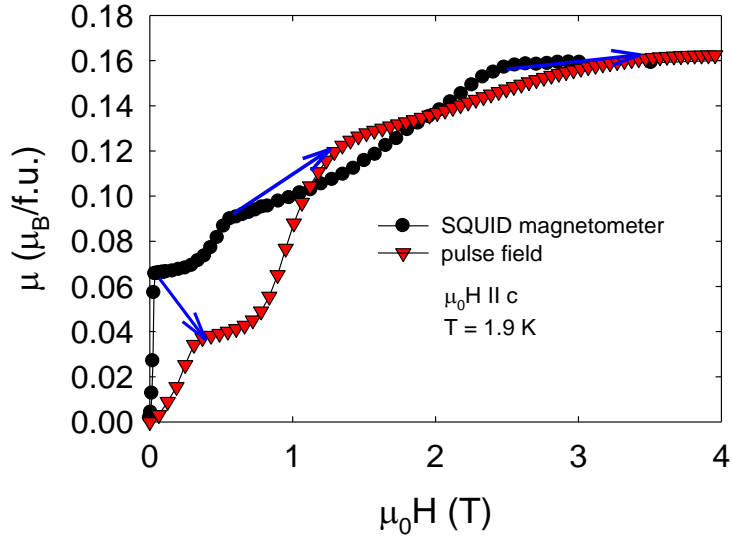

FIG. 13. The virgin magnetization curves measured in various magnetic field sweep rates. All jumps (plateaus) have been shifted to higher field in the case of pulse field magnetization experiment. The red (full line) arrows mark the shift of the 1/3 plateau. The blue (dashed) arrows mark the shift of the 1/2 jump.

The original 1/3 plateau has been shifted to higher magnetic field (from original field region $\mu_0 H_{C1} = 0.06$ T and $\mu_0 H_{C2} = 0.25$ T to $\mu_0 H_{C1} = 0.35$ T and $\mu_0 H_{C2} = 0.70$ T). In addition the original 1/3 plateau has not conserved the 1/3 characteristic and has been suppressed to the lower magnetic moment. The 1/2 jump has been shifted from the original $\mu_0 H_{C3} = 0.55$ T to $\mu_0 H_{C3} = 1.35$ T and saturation has been reached at a significantly higher magnetic field of $\approx$ 3.5 T. We have suggested a scenario of the kinetic effect on the magnetization of SmPd$_2$Al$_3$. There, we find large change of the magnetization process where even the height of the plateau changes. Thus, we have to consider that the ordered state has a rather complicated structure and various metastable states exist. In general, a plateau indicates a collinear structure (e.g., up-up-down structure, etc.), while gradual increase indicates a non-collinear structure (e.g., the Y shape structure). In Figs. 3, 4 and 5 we find that the temperature simply smears the structure, but in the Fig. 14, the spin structure at intermediate magnetic field seems different in the SQID measurement and the pulse field measurement. Unfortunately, at this moment, we cannot identify the structure, yet. Further detailed observations are expected.

**Conclusions**

Within the SmPd$_2$Al$_3$ study we have established the magnetic phase diagram on the basis of magnetization data. We have found rather complicated magnetic phase diagram where four different magnetic phases appear with pronounced hysteresis of two phases. We have detected the rather reduced saturated magnetic moment (0.16 μ$_B$/f.u.) then expected for Sm$^{3+}$ ion which is most probably given by the strong crystal field effect. Even applied high magnetic field of 60 T has not led to any significant increase of the saturated magnetization. Although the constructed magnetic phase diagram brings considerable progress in knowledge of the Sm magnetism in SmPd$_2$Al$_3$ compound the detail information regarding their magnetic structures were still lacking of.

Therefore we have performed a neutron diffraction experiment of SmPd$_2$Al$_3$ single crystal and we have successfully observed magnetic reflection (5/3 5/3 0) and its equivalents in the temperature interval 12.4 - 4.3 K which denotes the magnetic **k**-vector (1/3 1/3 0). Consequently SmPd$_2$Al$_3$ material can be considered as belonging to the group of magnetically frustrated systems. Based on our observations, we expect a triangular lattice antiferromagnet with weak Ising like anisotropy as the most suitable model for SmPd$_2$Al$_3$ compound.

The pulsed high magnetic field magnetization experiment surprisingly points to the influence of kinetic effect in magnetization process. The kinetic effect turns out to be the partially responsible effect for step like shape of magnetization curves at low temperatures where various rates of external field sweep lead to different metastable magnetic states.

On the basis of our investigation, the SmPd$_2$Al$_3$ compound represents a unique example of a complicated three dimensional phase diagram when not only the temperature and magnetic field are external variables but also the field sweep rate plays an important role. The rich magnetic phase diagram is given by a unique interplay of the magnetic frustration with kinetic effect of the sweeping magnetic field.

Although many features of the Sm magnetism in SmPd$_2$Al$_3$ have been conceived, there are still unresolved questions regarding the magnetic structures of the low temperature phases. Mainly the question and confirmation of the existence off-*c*-axis component of the magnetic moment seems to be most essential to understand the SmPd$_2$Al$_2$ physics well. Magnetic X-ray resonant scattering performed on Sm absorption edge, polarized neutron diffraction and fine torque magnetometry studies are planned to reveal relevant SmPd$_2$Al$_2$ magnetic features.

**Acknowledgements**

This work was supported by the Czech Science Foundation (Project # 202/09/1027) and the Charles University grant UNCE 11. Experiments performed in MLTL (see: http://mltl.eu/) were supported within the program of Czech Research Infrastructures (project # LM2011025). Neutron diffraction experiments in ILL Grenoble were performed within the project # LG11024 financed by the Ministry of Education of the Czech Republic. High-field magnetization measurements were supported by EuroMagNET under the EU contract n°228043.